\documentclass{ws-procs975x65}

\usepackage{amsfonts}
\usepackage{mathrsfs}
\usepackage{amsmath}

\def\be{\begin{equation}}       \def\ee{\end{equation}}
\def\beq{\begin{equation}}      \def\eeq{\end{equation}}
\def\bea{\begin{eqnarray}}      \def\eea{\end{eqnarray}}
\def\beqa{\begin{eqnarray}}      \def\eeqa{\end{eqnarray}}

\def\beqs{\begin{subequations}}
\def\eeqs{\end{subequations}}

\def\ifb{\textrm{fb}^{-1}}

\def\Si{\Sigma}

\def\f{\frac}

\def\nn{\nonumber}

\begin{document}

\title{Vector Bosons Signals of\\
Electroweak Symmetry Breaking\footnote{This contribution is an abbrievated version of ref. \cite{Du:2012vh}.}}

\author{R. Sekhar Chivukula$^*$, Elizabeth H. Simmons}
\address{Department of Physics and Astronomy,  Michigan State University\\
East Lansing, Michigan 48824, USA\\
$^*$ Speaker at Conference}

\author{Chun Du, Hong-Jian He, Yu-Ping Kuang, Bin Zhang}
\address{Center for High Energy Physics, Tsinghua University\\
 Beijing 100084, China}

\author{Neil D. Christensen}
\address{Pittsburgh Particle Physics, Astrophysics and Cosmology Center\\ Department of Physics and Astronomy\\ University of Pittsburgh,
Pittsburgh, PA 15260, USA}

\begin{abstract}
We study the physics potential of the 8\,TeV LHC (LHC-8) to discover signals of extended gauge models or extra dimensional
models whose low energy behavior is well represented by an
$SU(2)^2 \otimes U(1)$ electroweak gauge structure.
We find that with a combined integrated
luminosity of $40$\,fb$^{-1}$, the first new
Kaluza-Klein mode of the $W$ gauge boson can be discovered up to a
mass of about $400$\,GeV, when produced in association with a $Z$ boson.\end{abstract}


\bodymatter

\section{Introduction}

The ATLAS and CMS experiments at the LHC have now each collected over 20\,$\ifb$  of data at an 8\,TeV collision
energy. These data will enable the LHC to make incisive
tests of the predictions of many competing models of the origin of
electroweak symmetry breaking (EWSB), from the Standard Model (SM)
with a single Higgs boson, to models with multiple Higgs bosons, and
to so-called Higgsless models of the EWSB.
The Higgsless models \cite{Csaki:2003dt}
contain new spin-1 gauge bosons which
play a key role in EWSB by delaying unitarity violation
of longitudinal weak boson scattering up to a higher ultraviolet (UV)
scale \cite{SekharChivukula:2001hz}. Very recently, the effective UV completion of the minimal
three-site Higgsless model \cite{3site} was presented and studied
in \cite{Abe:2012fb} which showed that the latest LHC signals of a
Higgs-like state with mass around $125-126$\,GeV \cite{July4}
can be readily explained, in addition to the signals of new spin-1
gauge bosons studied in the present paper.

In this talk, we explore the physics potential of the LHC-8 to discover a
relatively light fermiophobic electroweak gauge boson $W_1$ with
mass $250-400$\,GeV, as predicted
by the minimal three-site moose model\,\cite{3site} and
its UV completion\,\cite{Abe:2012fb}.  Being fermiophobic or nearly so,
the $W_1$ state is allowed to be fairly light. More specifically,
the 5d models that incorporate ideally
\cite{SekharChivukula:2005xm} delocalized fermions
\cite{Cacciapaglia:2004rb,Foadi:2004ps}, in which the ordinary
fermions propagate appropriately in the compactified extra dimension
(or in deconstructed language, derive their weak properties from
more than one $SU(2)$ group in the extended electroweak sector
\cite{Chivukula:2005bn,Casalbuoni:2005rs}), yield phenomenologically
acceptable values for all $Z$-pole observables \cite{3site}. In this
case, the leading deviations from the SM appear in
multi-gauge-boson couplings, rather than the oblique parameters
$S$ and $T$. Ref.\,\cite{tri-3site} demonstrates that the LEP-II
constraints on the strength of the coupling of the $Z_0^{}$-$W_0^{}$-$W_0^{}$
vertex allow a $W_1$ mass as light as 250\,GeV,
where $W_0$ and $Z_0$ refer to the usual electroweak gauge bosons.

\section{The Model}

\begin{figure}[t]
\begin{center}
\includegraphics[scale=1.0]{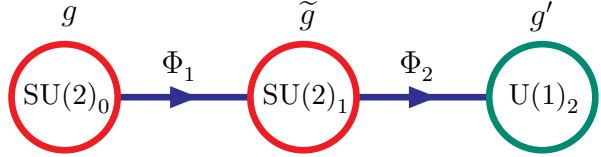}
\end{center}
\null\vspace{-1cm}
\caption{Moose diagram of the minimal linear moose model (MLMM)
with the gauge structure $SU(2)_0 \times SU(2)_1 \times U(1)_2$\,
as well as two independent link fields $\Phi_1^{}$ and $\Phi_2^{}$
for spontaneous symmetry breaking.
The relevant parameter space of phenomenological interest
is where the gauge couplings obey $\,g,g' \ll \tilde{g}$\,.}
\label{fig:models}
\end{figure}

We study the minimal deconstructed moose model at LHC-8
in a limit where its gauge sector is equivalent to the  ``three site model" \cite{3site}
or its UV completed ``minimal linear moose model" (MLMM)\,\cite{Abe:2012fb}.
Both the three site model and the MLMM are based on the gauge group
$\,SU(2)_0\otimes SU(2)_1\otimes U(1)_2$,
as depicted by Fig.\,\ref{fig:models} and its
gauge sector is the same as that of the BESS models \cite{Casalbuoni:1985kq,Casalbuoni:1996qt} 
or the hidden local symmetry model
\cite{Bando:1985ej,Bando:1985rf,Bando:1988ym,Bando:1988br,Harada:2003jx}.
The extended electroweak symmetry spontaneously breaks to electromagnetism
when the distinct Higgs link-fields $\Phi_1$ connecting $SU(2)_0$ to
$SU(2)_1$ and $\Phi_2$ connecting $SU(2)_1$ to $U(1)_2$ acquire
vacuum expectation values (VEVs) $\,f_{1}^{}$ and $\,f_{2}^{}$\,.\,
The weak scale $\,v \simeq 246$\,GeV is related to those VEVs via
$\,v^{-2} = f_1^{-2} + f_2^{-2}$\, %
and, for illustration, we take $\,f_1^{}=f_2^{}=\sqrt{2}v$\,.\,
Below the symmetry breaking scale, the gauge boson spectrum includes an extra
set of weak bosons $(W_1^{},\,Z_1^{})$,\, in addition to
the standard-model-like weak bosons $(W_0^{},\,Z_0^{})$ and the photon.
Furthermore, the scalar sector of the MLMM\,\cite{Abe:2012fb}
contains two neutral physical Higgs bosons
$(h^0,\,H^0)$, as well as the six would-be Goldstones eaten by
the corresponding gauge bosons $(W_0^{},\,Z_0^{})$ and $(W_1^{},\,Z_1^{})$.

One distinctive feature of the MLMM is that the unitarity of high-energy longitudinal
weak boson scattering is maintained jointly by the exchange of both the new
spin-1 weak bosons and the spin-0 Higgs bosons \cite{Abe:2012fb}.
This differs from either the SM (in which unitarity of longitudinal weak boson scattering
is ensured by the exchange of the Higgs boson alone)\,\cite{SMuni} or the conventional
Higgsless models (in which unitarity of longitudinal weak boson scattering
is ensured by the exchange of spin-1 new gauge bosons alone)\,\cite{SekharChivukula:2001hz}.
It has been shown \cite{tri-3site} that the scattering amplitudes
in such highly deconstructed models with only three sites
can accurately reproduce many aspects of the low-energy behavior
of 5d continuum theories.

The unitarity of the generic longitudinal scattering amplitude
of $\,W_{0}^{L}W_{0}^{L}\to W_{0}^{L}W_{0}^{L}$,\, in the presence of any numbers
of spin-1 new gauge bosons $V_k^{}\,(=W_k^{},Z_k^{})$
and spin-0 Higgs bosons $h_k$, was recently studied in Ref.\,\cite{Abe:2012fb}.
For the MLMM, tree-level unitarity implies sum rule\cite{Abe:2012fb},
%
\beqa
\label{eq:SR-MLMM}
G_{4W_0}^{} - \f{3M_{Z_0}^2}{4M_{W_0}^2}G_{W_0W_0Z_0}^2 \,=\,
\f{3M_{Z_1}^2}{4M_{W_0}^2}G_{W_0W_0Z_1}^2 +
\f{\,G_{W_0W_0h}^2\!+G_{W_0W_0H}^2\,}{4M_{W_0}^2} \,,
\eeqa
%
where the symbols $(h,\,H)$ denote the two mass-eigenstate Higgs bosons.
The sum rule illustrates how exchanging both the new spin-1 weak bosons $W_1/Z_1$ and the spin-0 Higgs bosons $h/H$
is required to ensure the
unitarity of longitudinal
weak boson scattering in the MLMM\cite{Abe:2012fb}.\footnote{
We also note that the $hWW$ and $hZZ$ couplings are generally suppressed\,\cite{Abe:2012fb}
relative to the SM values because of the VEV ratio $\,f_2^{}/f_1^{}=O(1)\,$
and the $h-H$ mixing.}

\section{Analysis of $\,{\bf W}_{\bf 1}^{\mathbf{\pm}}$ Detection at the LHC-8}

Extrapolating from our previous work \cite{PRD2008} at a
14 TeV LHC, we have found that the best process for detecting  $W_1^{}$ at LHC-8 is associated production,
$\,pp\rightarrow W_1^{}Z_0^{}\rightarrow W_0^{}Z_0^{}Z_0^{}
 \rightarrow jj \ell^+ \ell^-\ell^+ \ell^- $, 
where we select the $W_0^{}$ decays into dijets and the $Z_0^{}$ decays into
electron or muon pairs.

We have systematically computed all the major SM backgrounds
for the $jj4\ell$ final state,
including the irreducible backgrounds $\,pp\to W_0Z_0Z_0\to jj4\ell\,$ ($jj=qq'$)
without the contribution of $W_1$,\,
as well as the reducible backgrounds
 $\,pp\to ggZ_0Z_0\to jj4\ell$\,,\,
$\,pp\to Z_0Z_0Z_0\to jj4\ell\,$, and the SM $\,pp\to jj4\ell\,$
other than the above reducible backgrounds.

We performed parton level calculations at tree-level using two
different methods and two different gauges to check the consistency.
In one calculation, we used the helicity amplitude approach
\cite{helicity} to generate the signal and backgrounds. We also
calculated both the signal and background using CalcHEP
\cite{Pukhov:1999gg,Pukhov:2004ca}. For the signal calculation in
CalcHEP, we used FeynRules \cite{Christensen:2008py} to implement
the minimal Higgsless model \cite{Christensen:2009jx}. We found
satisfactory agreement between these two approaches and between both
unitary and 't\,Hooft-Feynman gauge. We used a scale of
$\,\sqrt{\hat s}$\, for the strong coupling in the backgrounds and
$\,\sqrt{\hat s}/2$\, for the CTEQ6L \cite{cteq6} parton
distribution functions.  We included both the first and second
generation quarks in the protons and jets, and both electrons and
muons in the final-state leptons.

In our calculations, we impose basic acceptance cuts,
\begin{eqnarray}
 p_{T \ell}^{} > 10\,{\rm GeV},  ~~~~~  |\eta_\ell^{}|<2.5 \,,
\nn\\[0.9mm]
 p_{T j}^{} > 15\,{\rm GeV},  ~~~~~  |\eta_j^{}|<4.5 \,,
\end{eqnarray}
and also a reconstruction cut for identifying $W_0$ bosons that
decay to dijets,
\begin{eqnarray}
 M_{jj}^{} = 80 \pm 15\,{\rm GeV} \,.
\end{eqnarray}

We further analyzed the distributions of the dijet opening-angle
$\,\Delta R(jj)$\, in the decays of $\,W_0\to jj$\, for both the
signal and SM background events. We find that the signal events are peaked in the
small opening-angle region around $\,\Delta R(jj)\sim 0.6$\,,\,
while the SM backgrounds tend to populate the range of larger
opening angles, with a broad bump around $\,\Delta
R(jj)=1.5-3.3$\,. In order to sufficiently suppress the SM
backgrounds, we find the following opening-angle
cut\footnote{These are somewhat weaker than the cut of %
$\,\Delta R(jj)< 1.5\,$ imposed in \cite{PRD2008}.} to be very
effective \cite{j-separation},
\beqa
\label{eq:DR-cut1}
\Delta R(jj) ~<~ 1.6\,.
\eeqa
At the LHC-8, we note that the above cut reduces the signal events
by only $10-15\%$, but removes about $72-80\%$ of the SM backgrounds.

\begin{figure}[t]
  \centering
   \includegraphics[width=8cm]{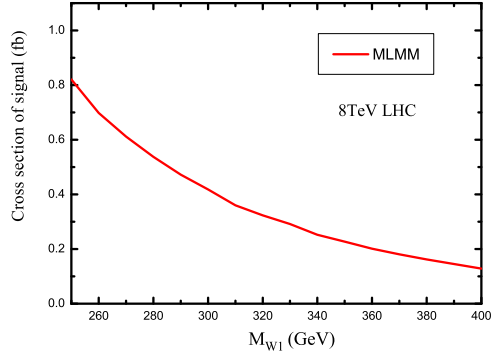}
\null\vspace{-5mm}
  \caption{Predicted signal cross section for
 $\,pp\to W_1Z_0\to W_0Z_0Z_0\to jj4\ell\,$ as a function of the
 $W_1^{}$ mass in the MLMM after all cuts at the LHC-8.}
 \label{Fig:3}
\end{figure}

In Fig.\,\ref{Fig:3}, we display the predicted total signal cross section
for the process $\,pp\to W_0Z_0Z_0\to jj4\ell\,$
after all cuts at the LHC-8 have been imposed; this is shown as a function of the $W_1^{}$
mass for the range $\,250-400$\,GeV.\footnote{Here, we define the signal region to include all events satisfying the condition,
$ M({Z_0jj}) ~= ~M_{W_1}^{} \pm 20 \,\text{GeV}$~.}

In Fig.\,\ref{Fig:5}, we display the required integrated
luminosities for detecting the $W_1^\pm$ signal at the $3\sigma$ and
$5\sigma$ levels as a function of the $W_1^\pm$ mass $M_{W_1}^{}$.  We see the LHC-8 should be
able to observe the $W_1^\pm$ gauge bosons
of the minimal linear moose model studied up to masses of order 400 GeV. We look forward to seeing the results.

\begin{figure}[t]
\hspace*{-9mm}%
  \centering
   \includegraphics[width=8cm]{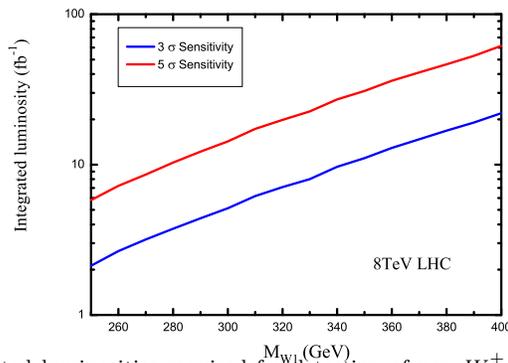}
   \vspace*{-7mm}
  \caption{Integrated luminosities required for detection of
  new $W_1^\pm$ gauge bosons at the $3\sigma$ level in the MLMM (lower blue curve),
  and at the $5\sigma$ level (upper red curve) as a function of the $W_1^{}$
  mass, at the LHC-8.} %
\label{Fig:5} %
\end{figure}

\vspace*{6mm} %
\noindent %
{\bf Acknowledgments}
\\[1mm]
This research was supported by the NSF of China (grants 11275101, 10625522,
10635030, 11135003, 11075086) and the National Basic Research
Program of China (grant 2010CB833000); by the U.S. NSF under Grants
PHY-0854889 and PHY-0705682; and by the University of Pittsburgh
Particle Physics, Astrophysics, and Cosmology Center. HJH thanks
CERN Theory Division for hospitality.


\end{document}